# MaxwellNet: Physics-driven deep neural network training based on Maxwell's equations

Joowon Lim [1*] and Demetri Psaltis [1]

[1] École Polytechnique Fédérale de Lausanne, Optics Laboratory, CH-1015 Lausanne, Switzerland.

## Abstract

Maxwell's equations govern light propagation and its interaction with matter. Therefore, the solution of Maxwell's equations using computational electromagnetic simulations plays a critical role in understanding light-matter interaction and designing optical elements. Such simulations are often time-consuming and recent activities have been described to replace or supplement them with trained deep neural networks (DNNs). Such DNNs typically require extensive, computationally demanding simulations using conventional electromagnetic solvers to compose the training dataset. In this paper, we present a novel scheme to train a DNN that solves Maxwell's equations speedily and accurately without relying on other computational electromagnetic solvers. Our approach is to train a DNN using the residual of Maxwell's equations as the physics-driven loss function for a network that finds the electric field given the spatial distribution of the material property. We demonstrate it by training a single network that simultaneously finds multiple solutions of various aspheric micro-lenses. Furthermore, we exploit the speed of this network in a novel inverse design scheme to design a micro-lens that maximizes a desired merit function. We believe that our approach opens up a novel way for light simulation and optical design of photonics devices.

**Keywords:** Maxwell's equation, Deep learning, Computational electromagnetics, Finite difference method, Inverse design;

---

*Correspondence: J. Lim (joowon.lim@epfl.ch)

Maxwell's equations describe how light propagates and interacts with materials and account for basic light properties such as refraction, diffraction, scattering, and so on[1–3]. The computational electromagnetic simulations which numerically solve Maxwell's equations are essential in understanding light-matter interactions near or below the scale of wavelength and designing optical elements at such scales[4–7].

Recently, there is active research aimed at replacing the time-consuming computational electromagnetic simulators with deep neural networks (DNNs)[8] trained to predict optical properties when given the structural parameters[9–13]. However, they generally require datasets of paired input-output examples to train the networks. As a result, these networks require numerous simulations to construct the training datasets, which is very laborious. Instead of directly predicting the optical properties, DNNs have been also used to speed up the convergence of the Generalized Minimal Residual algorithm, but the output from the network should be followed by iterative calculations to find the electromagnetic field[14].

Another branch of active research involves DNNs to solve partial differential equations in an indirectly supervised manner[15–18]. The key idea is to use the residual of the target differential equation as a loss function and train the network parameters to find the solution. To the best of our knowledge, this approach has not been applied to solve Maxwell's equations for inhomogeneous dielectric material distribution.

In this contribution, we propose to use the residual of Maxwell's equations as a physics-driven loss function to train a DNN that finds the electric field, given the material property distribution as its input. We demonstrate our approach by training a network that finds the electric field distributions for various aspheric micro-lenses. Furthermore, using the DNN trained for various micro-lenses, we demonstrate that we can inverse design a micro-lens that maximizes the light focusing at a target point. To do so, another DNN is introduced to encode and represent the shape of lenses at a low dimensional *latent* space[19,20], and we optimize the shape in this latent space.

Maxwell's equations can be written as follows,

$$\nabla \times \nabla \times \boldsymbol{E}(\boldsymbol{r}) - k_0^2 \varepsilon_r(\boldsymbol{r})\boldsymbol{E}(\boldsymbol{r}) = 0, \qquad (1)$$

for linear, non-magnetic, and isotropic materials without electric and magnetic current densities. $\boldsymbol{E}$ represents the electric field vector, $k_0 = 2\pi/\lambda$ is the wavevector given a wavelength in the air, $\lambda$, and $\varepsilon_r(\boldsymbol{r})$ represents the relative electric permittivity distribution which is related to the refractive index (RI) distribution, $n(\boldsymbol{r})$, as follows $\varepsilon_r(\boldsymbol{r}) = n(\boldsymbol{r})^2$. Given the electric field and relative permittivity distributions, Eq. (1) can be used as a strong physics-based metric to evaluate the validity of the electric field. We carried out a simple example where we calculated the electric field for the same material property distribution using two methods that are expected to have different levels of accuracy, to visualize what the physics-driven loss function looks like. We present this in Fig. 1a. The illumination is a plane wave propagating in $z$ and polarized along the y-axis and the sample is a 2D plano-convex spheric microlens. The physics-driven loss, Eq. (1), was calculated on the Yee grid[21] with the high order approximation of the gradient operator. The Born field was calculated assuming the first-order Born approximation which assumes the electric field within the sample is the same as the incident electric field. Please refer to the supplementary material for detailed information. As the Born approximation is not valid in this case, we can see that the subtraction of two terms in Eq. (1) does not cancel each other, and leaves signals in the residual map within the sample. The ground truth in Fig. 1a was acquired using a commercial finite element method solver, COMSOL. We can confirm that the physics-driven loss calculated on the ground truth solution shows negligible discrepancy in the residual map.

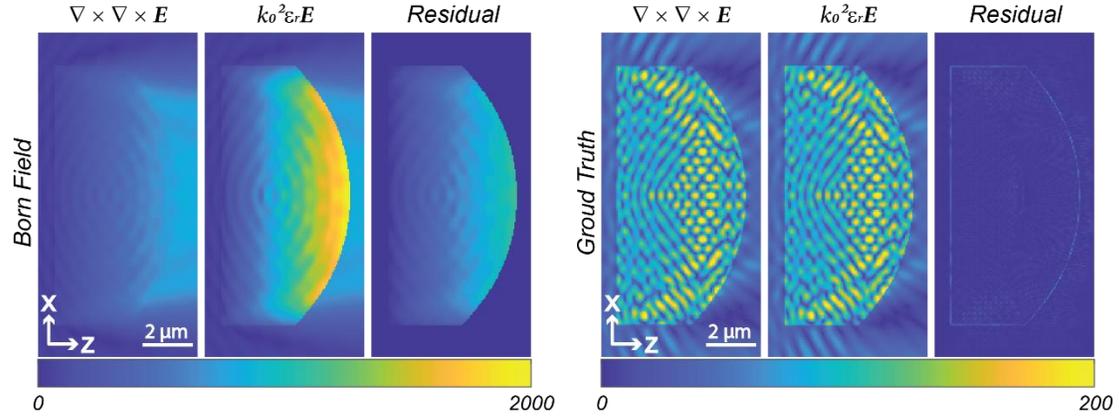

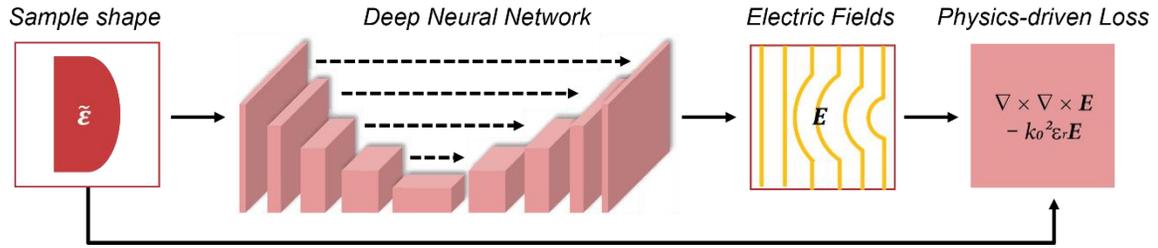

Figure 1. Examples of the physics-driven loss and description of the main idea. (a) Residuals of Maxwell's equations calculated on two electric fields: the electric field calculated under the first-order Born approximation (Born field) and the ground truth solution acquired using COMSOL (Ground truth). (b) Normalized electric permittivity values are given as input to the network, and the residual of Maxwell'sequations is calculated on the output electric field distributions, which is used as a physics-driven loss function to train the network parameters.

We use Eq. (1) as a physics-driven loss function to train a DNN. Since Maxwell's equations are true for all electromagnetic fields, we expect to converge to a network that predicts the field accurately by minimizing this loss function even in the absence of the ground truth field. As shown in Fig. 1b, the output of the network is the 3D distribution of the electric field vector, with $\varepsilon_r(r)$ as the input. In fact, the normalized relative permittivity distributions, $\tilde{\varepsilon}(r) = (\varepsilon(r) - \varepsilon_{min})/(\varepsilon_{max} - \varepsilon_{min})$ is used where $\varepsilon_{max}$ and $\varepsilon_{min}$ denote the maximum and minimum relative permittivity values. We refer to this network as MaxwellNet and this type of learning that takes place in MaxwellNet as *indirect* training. An alternative approach to finding a DNN that can calculate the fields given the material distribution consists of a first step where highly accurate electromagnetic simulators are used to generate a database of input-output pairs followed by a DNN that uses this database to learn to predict the field. We would refer to this as *direct* training since the network has access to the desired output for each input in the training set. In MaxwellNet, we only provide the inputs yet the network learns to *indirectly* infer the correct solution (field) due to the physics-driven loss. Indirect training can be done without access to a large database of input-output pairs which are either experimentally or computationally generated. This can be a big advantage in cases where measuring the 3D field distribution experimentally is not possible or calculating it computationally not practical.

In the following section, we use the MaxwellNet to design lenses to accomplish a specific task. We do this by using examples of multiple lens designs as a training set to search for new lenses. Therefore, we would like the MaxwellNet to be trained to predict the correct field for various index distributions in a class of aspheric lens. Unlike the spherical lens whose shape is described by one parameter (radius of curvature), the sag of an aspheric lens is defined by multiple parameters as follows when we neglect higher order terms[22],

$$z(r) = -\frac{r^2}{R(1+\sqrt{1-\frac{(1+\kappa)r^2}{R^2}})}, \qquad (2)$$

where $r$ is the radial coordinate, $R$ represents the radius of curvature, and $\kappa$ denotes the conic constant. We generated multiple aspheric lenses in 2D by changing these parameters (assuming no variation of $\varepsilon(r)$ along the $y$-axis). Please refer to the supplementary material for details. We separately trained two networks for transverse electric (TE) and transverse magnetic (TM) modes using this dataset.

Figure 2 demonstrates the results for four different examples, two each, for the TE and TM modes. We can see the changes in the electric field distributions for the different lens shapes, and the results from the network show great consistency with the ground truth solutions. For the TM mode where two polarized fields couple with each other, the incident light is polarized along the $x$-axis, however, the resulting fields are polarized in both $x$ and $z$ axes. The networks did not access any of the target solutions but were trained only using the physics-driven loss for both TE and TM modes.

In a recent paper[28], the authors trained a computational fluid dynamic simulator and used it to optimize the shape of a car to minimize the drag on it. In a similar way, thanks to the physics-driven loss, we can train a light-scattering simulator in the *indirectly* supervised manner, and use it to design an optical element to maximize a certain figure of merit (FOM). We demonstrate it using the trained MaxwellNet for various aspheric lenses to design an aspheric lens to focus light at the desired point. We should point out that this lens design problem is not something that can be done with standard ray optics tools since the dimensions are only 10λ and wave analysis must be used. The full-wave equation solvers that might be used to optimize the shape of the lens is the finite difference time domain (FDTD) method to calculate forward and backward propagations[4]. By exploiting the fast computation speed of MaxwellNet (less than 10 ms for the TE mode calculation), we can accelerate the inverse design process by replacing FDTD with MaxwellNet.

Since our goal is to design the shape of the lens, we introduce a *second* DNN, DeepSDF, that is trained to classify each pixel as either being the lens material or air. This is depicted in Fig. 3. Each sample in the training set consisting of aspheric lenses is assigned a latent vector, $v_i \in \mathbb{R}^N$ where the subscript $i$ denotes the index of sample. The network takes an input vector which is the concatenation of a latent vector, $v_i$, and a coordinate vector, $(x, z) \in \mathbb{R}^2$, and returns a scalar value, $s \in \mathbb{R}$. The output value is trained to be either $-0.5$ or $+0.5$ to represent the material at the point, $(x, z)$, for the $i$-th sample. The pixel value at a certain position, $(x, z)$, can be either the background or lens material depending on the lens shape which is determined by Eq. (2). DeepSDF is trained in a supervised way using the aspheric lens dataset, and each latent vector is also updated to encode the shape of the assigned lens. Please refer to the supplementary material for additional details. Once the network is trained, each latent vector encodes the shape of the assigned lens and returns it when given to the network. We selected two latent vectors assigned to two lenses in the training dataset and reconstructed the shapes of corresponding lenses as shown in Fig. 3b. We only trained half of the lenses as they are $x$-symmetric. Comparing with the ground truth lens shapes, we can confirm that the latent vectors along with DeepSDF can represent the lens shape with high accuracy. The significance of the latent space representation is that it is a compressed version of the complete shape of the lens and therefore it becomes computationally efficient to carry out the optimization in the latent space.

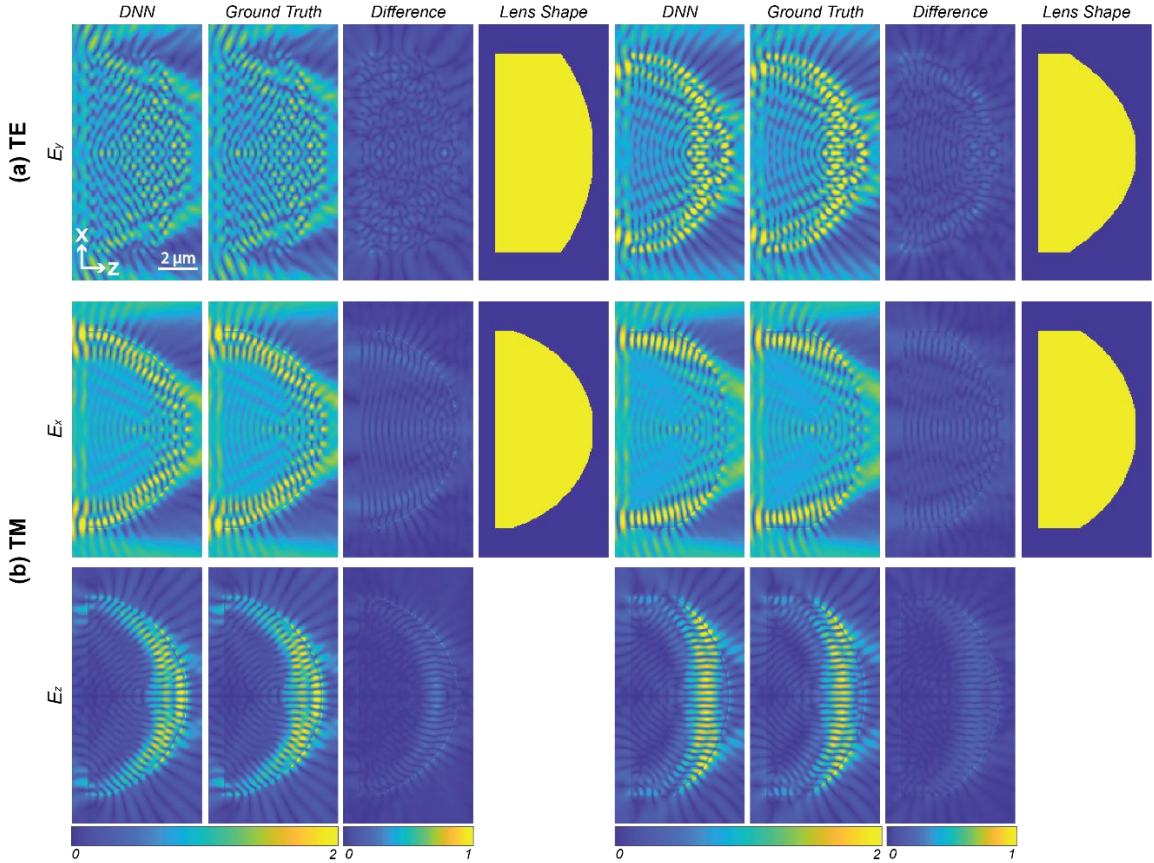

Figure 2. Results for four different examples, two each, for the (a) TE and (b) TM modes. The outputs of the networks are presented and compared with the ground truth solutions. For the TE mode, the output field only exists in the same polarization of the illumination ($E_y$). By contrast, the TM mode results present the fields not only in the incident polarization ($E_x$) but also in $z$ polarization ($E_z$) due to the polarization coupling.

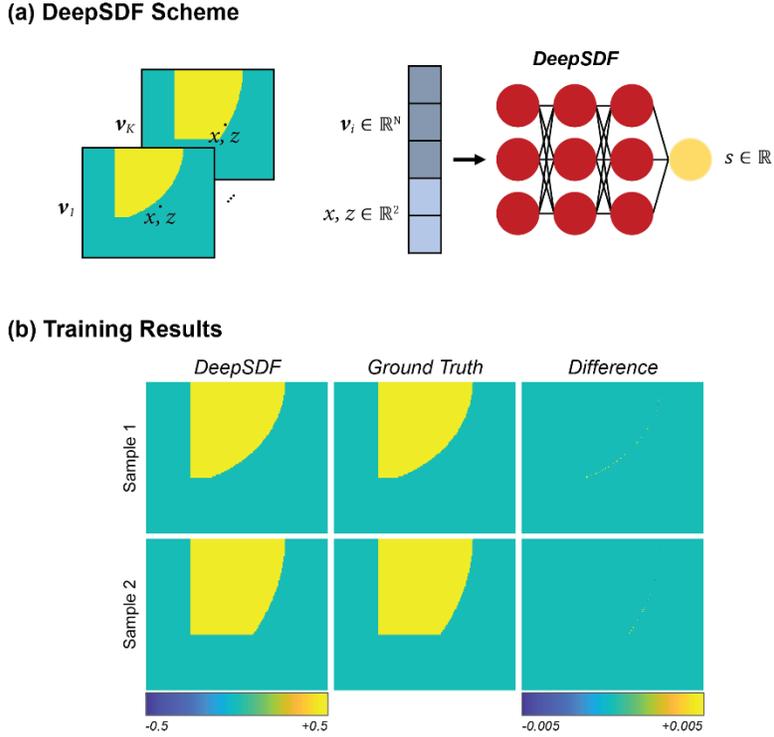

Figure 3. DeepSDF training. (a) The input to DeepSDF is the concatenated vector of a latent vector, $v_i$, and a position vector, $(x,z)$. The network is trained to give a scalar value as output to represent whether the position is filled with material (+0.5) or not (-0.5). (b) Two different lens shapes were reconstructed by giving two latent vectors to DeepSDF and compared with the corresponding target shapes.

Here, we propose a novel inverse design scheme by combining the MaxwellNet and DeepSDF. We demonstrate it by designing a micro-lens to maximize the light intensity at a target position. The inverse design scheme is described in Fig. 4a. Given a latent vector, $v_i$, along with the position values, DeepSDF produces the corresponding lens shape, and it serves as an input to MaxwellNet (TE mode) to calculate the corresponding electric field distribution. We can efficiently calculate the electric field outside of the sample using the homogeneous medium Green's function (Details are in supplementary material), and define a figure of merit (FOM) function as the intensity at the target focal point. Usual inverse design approaches maximize the FOM function by taking the derivative with respect to the sample shape, $\frac{\partial FOM}{\partial s}$, and the solution is usually not discrete nor manufacturable. Unlike the conventional inverse design approaches, the proposed scheme maximizes the FOM with respect to the latent vector, $\frac{\partial FOM}{\partial v}$. In other words, since DeepSDF learned to represent the various lens shapes when given the latent vectors which encode the shapes, we can optimize the lens design by finding a latent vector that maximizes the FOM.

Fig. 4b shows an example of designing a micro-lens to focus light at 8 $\mu m$ from the center of the lens. The initial lens design is a spheric lens, and it forms its focus after 12 $\mu m$ from the center of the lens. With the spheric lens as an initial solution, we optimized the latent vector to maximize the intensity pixel value at 8 $\mu m$, we can confirm that the optimized aspheric lens focuses the light at the desired target location. To validate the result, we ran a COMSOL simulation for the optimized aspheric lens and compared the intensity profiles at the central line. We can see in the intensity profiles from the MaxwellNet and COMSOL results show great consistency.

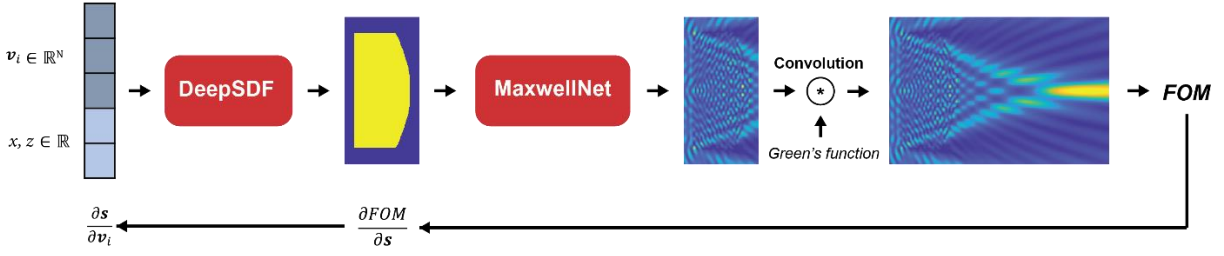

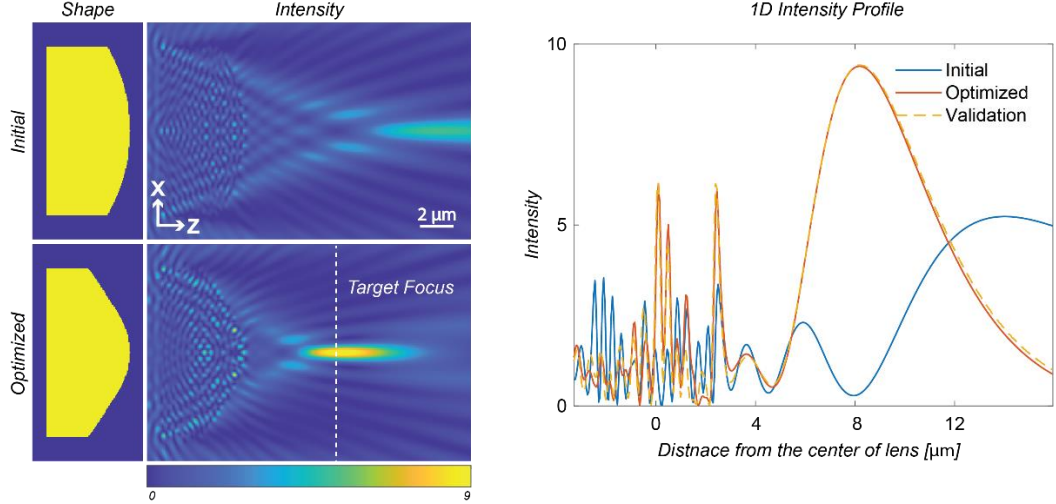

Figure 4. Description of the inverse design idea. (a) The input to DeepSDF is the concatenated vector of a latent vector, $v_i$, and a position vector, $(x, z)$, which produces a micro-lens shape, and it is followed by MaxwellNet to predict the electric field distribution. Using the Green's function, we can expand the computation domain. We optimize the lens shape in the latent space of DeepSDF to maximize the figure of merit function. (b) Starting from the initial spheric lens, we inverse-designed an aspheric lens which maximizes the light intensity value at 8 $\mu m$. The 1D profile plots at the center are presented for the initial and inverse-designed solutions. We further performed a COMSOL simulation on the inverse-designed solution to validate.

To summarize, we have described a physics-driven loss to train a DNN. The main idea is to penalize the residual of Maxwell's equations as a loss function to train a network that predicts electric field distributions when given the permittivity distribution of a sample. We applied the idea to simultaneously training the network parameters to predict the electric field distributions of a collection of aspheric lenses. The outputs from the network show great consistency with the ground truth solutions, we can therefore use the trained network, MaxwellNet, as an extremely rapid light scattering simulator for the trained lenses and similar ones. Here, we emphasize that the training process does not require any access to the target solutions but only uses the physics-driven loss based on Maxwell's equations. Furthermore, we proposed a novel inverse design scheme using MaxwellNet as the forward simulator. In conventional inverse design schemes based on the adjoint method[4], the calculated gradient values are continuous while we can only use specific materials whose permittivity values are discrete and distinct in most cases. It, therefore, requires additional steps to be integrated during the optimization process to guarantee that the final solution is manufacturable. To circumvent this problem, we trained DeepSDF along with the latent vectors to encode and represent the different lens shapes. By doing so, we can update the solution which maximizes the FOM function in the latent vector space rather than the shape space itself. We demonstrated it by designing a micro aspheric lens to maximize the light intensity at a desired focal point and validated the final lens design using COMSOL. Here, we have demonstrated the proposed scheme only for the microlens design, however, we believe that it can be further extended to other types of photonic design applications and pave a new way of inverse design.

## MATERIALS AND METHODS

*Network training:* For the MaxwellNet training, we discretized the physics-driven loss, Eq. (1), on the Yee grid, and approximated the gradient operators using the high order finite difference method. In order to have a unique solution, we further needed to incorporate the Sommerfeld's radiation condition on the scattered field by imposing the perfectly matching layers (PMLs)[23,24]. The implementation details are explained in the supplementary material. Using the symmetry of the samples and the corresponding electric field distributions, we could reduce the computation domain by half. We implemented the networks using PyTorch (1.7.1)[25] and trained them on graphic processing units (GPU, NVIDIA V100). After training, the other computations were performed on a desktop computer (Intel Core i7-6700 CPU, 3.4 GHz, 32 GB RAM) with a GPU (GeForce GTX 1070). For the other details of the network training and structures, please refer to the supplementary material. For every simulation present in the paper, the wavelength was set as $\lambda = 1\ \mu m$ with the spatial discretization of $50\ nm$. The RI values of the material and background were set as 1.53 and 1.0. The pixel numbers along the $x$ and $z$ axes were 320 and 192, respectively, but we reduced the pixel number along the $x$-axis by half using the symmetry.

*Inverse design:* During the inverse design process, all the network parameters of MaxwellNet and DeepSDF were fixed and we only updated the latent vector. The Adam optimizer[26] was used to update the latent vector with a learning rate of 0.01 and 0.9 was used for both coefficients for computing running averages of gradient and its square. The inverse design process was run for 200 iterations and the latent vector which showed the maximal FOM was kept as the final design.

*COMSOL simulation:* In order to evaluate the output electric fields from MaxwellNet, we performed a full-wave simulation in 2D using a commercial finite element method solver, COMSOL Multiphysics 5.4. The PMLs were used for every COMSOL simulation presented in this work. The resulting outputs from COMSOL were sampled at the grid points to be directly compared with the outputs of MaxwellNet. The comparisons were performed in MATLAB R2020b (MathWorks Inc., Natick, MA, USA) on a desktop computer (Intel Core i7-6700 CPU, 3.4 GHz, 32 GB RAM).

## ACKNOWLEDGMENTS

This work was funded at EPFL by Swiss National Science Foundation (514481).

## CONFLICT OF INTERESTS

The authors declare no conflict of interest.

## CONTRIBUTIONS

J.L. carried out the modeling and computations and D.P. supervised the project. All authors contributed to the discussion and wrote the manuscript.